\documentclass[twocolumn,showpacs,preprintnumbers,amsmath,amssymb,aps,prl]{revtex4}

\usepackage{color}
\definecolor{Blue}{rgb}{0.3,0.3,0.9}
\definecolor{Green}{rgb}{0.3,0.9,0.3}

\usepackage[dvips]{graphicx}
\usepackage{amsmath}
\usepackage{epsfig}

\newcommand{\ket}[1]{\left | \, #1 \right \rangle}
\newcommand{\bra}[1]{\left \langle #1 \, \right |}
\newcommand{\av}[1]{\langle #1\rangle}
\begin{document}

\title{Tailored particle current in an optical lattice by a weak time-symmetric harmonic potential}

\author{Julio Santos, Rafael A. Molina, Juan Ortigoso and Mirta Rodr\'{\i}guez}

\affiliation{Instituto de Estructura de la Materia, CSIC, Serrano 121-123, 28006 Madrid, Spain.}

\date{June 8, 2011}
\begin{abstract}
Quantum ratchets exhibit asymptotic currents when driven by a time-periodic potential of zero mean if the proper spatio-temporal symmetries are broken. There has been recent debate on whether directed currents may arise for potentials which do not break these symmetries.
We show here that, in the presence of degeneracies in the quasienergy spectrum, long-lasting directed currents can be induced, even if the time reversal symmetry is not broken. Our model can be realized with ultracold atoms in optical lattices in the tight-binding regime, and we show that the time scale of the average current can be controlled by extremely weak fields.
\end{abstract}
\pacs{05.60.Gg, 03.75.Kk, 37.10.Jk, 67.85.Hj}

\maketitle

Brownian motors or ratchets are spatially periodic systems with noise and/or
dissipation in which a directed current of particles can emerge from
an unbiased zero-mean external force \cite{Reimann_2002,Hanggi_Marchesoni_2009}. Models for biological engines that
transform chemical energy into unidirectional mechanical motion behave as
Brownian motors \cite{Julicher_1997}. Extensive studies of the ratchet effect in classical systems \cite{Schanz2001}
stated the relation between symmetry breaking potentials and the existence of the asymptotic current \cite{Flach_2000}. 
For a system driven by a flashing potential of the form $V(x,t)=V_x(x)V_t(t)$ with $V_t(t)=V_t(t+T)$ of zero mean and $V_x(x)=V_x(x+L_x)$
there are up to four different symmetries in the classical system that must be broken in order to generate an asymptotic current \cite{Denisov2007}. 
A ratchet current arises if one breaks the relevant spatio-temporal symmetries, here denoted by $S_x$, and the time-reversal symmetry 
$S_t: (x,p,t) \rightarrow (x,-p,-t+2t_s)$. Lately there has been an increasing interest in the coherent ratchet effect in Hamiltonian quantum systems
\cite{Reimann_Grifoni_Hanggi_2002}. It has been shown that the same symmetry requirements
apply to them \cite{Denisov2007}, i.e. if the Hamiltonian preserves any of the symmetries,
no asymptotic current is possible.

Experimentally, directed current generation was first studied in solid state devices, quantum dots and Josephson
junctions \cite{Solidstate}. More recently, the precise control achievable in cold atom experiments opened up the possibility of realizing directed atomic currents for Hamiltonian systems 
with controllable or no dissipation in the time scale of the measurements \cite{Schiavoni2003,Gommers2005,Jones_2007,QAM}.
Recently, a very clean realization of a coherent quantum ratchet was experimentally demonstrated in
a Bose-Einstein condensate exposed to a sawtooth potential realized with an optical lattice which was periodically modulated in time \cite{Salger_2009}.
Directed transport of atoms was observed when the driving lattice potential broke the spatio-temporal symmetries. 
The current oscillations and the dependence of the current on the initial time and the resonant frequencies \cite{Res_2007} were measured, demonstrating the quantum character of the ratchet.

% Problem of the non-asymptotic current

Although the generation of an asymptotic directed current needs the breaking of the symmetries $S_x$ and $S_t$ simultaneously for unbiased potentials,
there has been a recent discussion on the possibility of obtaining long-lasting directed currents without it \cite{ Hanggi2009,Sols,comment,reply,Moskalets}.
Many-body effects \cite{Hanggi2009,Moskalets} with the proper choice of the initial state \cite{Hanggi2009} or an accidental degeneracy in the quasienergy spectrum \cite{reply} may result in a directed current without breaking the time-reversal symmetry.  
In contrast to previous works we show here that one can exploit a quasi-degeneracy, present for a wide range of parameters, in the quasienergy spectrum in order to generate a long-lasting directed average current in a weakly driven system where we can achieve full control over its magnitude and time scale. \\
Previous work on quantum accelerators \cite{QAM} has shown that in the presence of quantum resonances one can obtain large currents without breaking the time-reversal symmetry using a delta-kicked potential in time. 
%Quantum accelerator modes result of wave packets trapped in a stable island of a classical map, which decay by tunneling out to the chaotic sea.
Essentially, for particular values of the Hamiltonian parameters  the spectrum of the Floquet operator becomes continuous due to quantum resonances.  
%Particular initial states show 
Under such circumstances one can obtain a linear increase in momentum with time which has been claimed as a true ratchet effect driven by resonances instead of noise \cite{QAM2}. 
%Further work \cite{QAM3}, suggested that unbounded momentum generation is generic and it can be presented in fully chaotic as well as in nearly-integrable systems.  
However,  there is no formal proof that the dynamics show unbounded acceleration for times longer that those that are computed  \cite{QAM}. Nonetheless, a significant difference between quantum accelerator ratchets and our system is the existence of a constant component in the delta potential.

%It is known that the inclusion of a dc-component in the driving external field may lead to a directed transport against a constant bias even in the Hamiltonian limit \cite{QAM5}.

One useful way of treating time-periodic quantum Hamiltonians, $H(t)=H(t+T)$, is the Floquet formalism \cite{Floquet}. The cyclic states $\ket{ \phi_j (t+T)}=e^{-i \varepsilon_jT}\ket{ \phi_j (t)}$ are the eigenstates of the evolution operator for one period while the quasienergies $\varepsilon_j$ are the eigenvalues.
The solution to the time-dependent Schr\"odinger equation
$H(t) \ket{\psi(t)}=i\hbar{\partial} \ket{\psi(t)}/{\partial t}$ can be spanned in the cyclic eigenbasis ($\hbar=1$)
\begin{eqnarray}
&& |\psi(t)\rangle=\sum_{j} e^{-i \varepsilon_j t} c_j \ket{\phi_{j}(t)}
\label{eqjulioq} 
\end{eqnarray}
where $c_j=\langle\phi_{j}(0)|\psi(0)\rangle$.
% \cite{Floquet}
The average current generated during $n$ cycles is given by $\mathcal{I}(t=nT)=\frac{1}{n} \sum_{m=1}^n \mathcal{I}_m $ with
\begin{equation}
 \mathcal{I}_m=\frac{1}{T} \int_{(m-1) T}^{mT}  \langle \psi(t)|p|\psi(t)\rangle dt,\label{eq:Im}
\end{equation}
where $p$ is the
momentum operator.
Note that due to the periodicity of the cyclic states the average current during $n$ cycles can be simplified in terms of integrals of the cyclic states during the first period
\begin{eqnarray}
\mathcal{I}(n T) &=&\frac{1}{n}\sum_{j,j'} c_j c^{*}_{j'} \av{p}_{j j'} \frac{1-e^{-inT(\varepsilon_j-\varepsilon_{j'})}}{1-e^{iT(\varepsilon_j-\varepsilon_{j'})}},  \label{eq:JnT}  
%\\ \av{p}_{j j'} &\equiv& \frac{1}{T} \int_0^T \bra{\phi_j (t)} p \ket{\phi_{j'} (t) }e^{-it(\varepsilon_j-\varepsilon_j')}dt.
\end{eqnarray}
which is valid for a discrete Floquet spectrum and where $\ \av{p}_{j j'} := \frac{1}{T} \int_0^T \bra{\phi_j (t)} p \ket{\phi_{j'} (t) }e^{-it(\varepsilon_{j'}-\varepsilon_j)}dt$. 
Our model, that represents well optical lattices, has by construction a pure point spectrum. In the limit of an infinite number of states the single-band tight-binding model remains integrable \cite{spectrum1}. However, if more bands are added the spectrum could become absolutely continuous (in resonance) or singular continuous. Our guess, based on our previous experience in related problems \cite{spectrum2}, is that except at very long times even if the spectrum is singular continuous the time evolution will be similar to that given by a pure point spectrum.
In general, a sum of oscillatory off-diagonal terms with arbitrary exponents decays
rapidly \cite{decay} and for long times only the diagonal terms in Eq.~(\ref{eq:JnT}) remain.
If both $S_x$ and $S_t$ are broken, the cyclic eigenstates desymmetrise and carry net momentum, i.e.
$ \av{p}_{jj} \neq 0 $ \cite{Denisov2007}. In such case, the asymptotic average current at $n\rightarrow \infty$ is nonzero $\mathcal{I}(\infty)=\sum_{j} |c_j|^2 \av{p}_{jj}$. Correspondingly, if either of the relevant symmetries is not broken $ \av{p}_{jj} = 0 $ and thus $\mathcal{I}(\infty)=0$.
Note, however, that the off-diagonal terms in Eq.~(\ref{eq:JnT}) become relevant if the initial state projects mainly into degenerate or quasidegenerate cyclic states with
$\av{p}_{j j'}\neq 0$. 
If one induces a resonance between the proper  quasienergy states at low driving, 
the average current  contains only a small number of terms in the sum and the exponents can be very small, leading to very slow oscillations whose period can be fitted by tunning the driving. 
In order to maximise the average current one should then optimise both the projection into the initial states $c_j c^{*}_{j'}$ and the $\av{p}_{j j'}$. 
%We illustrate this here and show that one can obtain measurable long-lasting currents by tuning the parameters of the model in an appropriate way. \\
We illustrate this here and show that it is possible to populate a high average momentum superposition of cyclic eigenstates for times which can be tuned up to the lifetime of the experiment.  \\
%%%%%%%%%%
\begin{figure}
%\begin{tabular}{c}
%\epsfig{file= fig1.eps,width=0.9\linewidth,clip=} 
\epsfig{file= 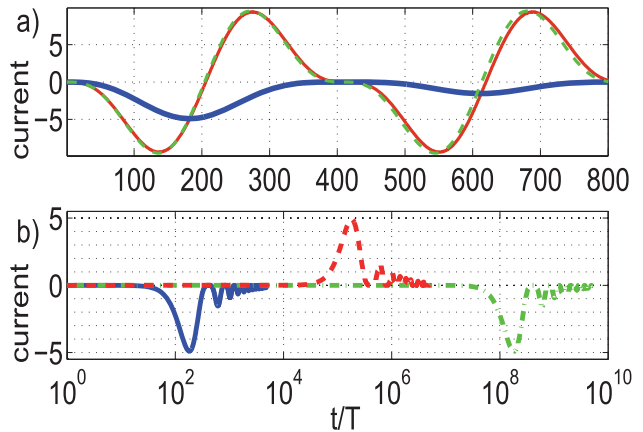,width=\linewidth,clip=}
%\end{tabular}
\caption{ (color online) a) Thick solid line shows the average current, $\mathcal{I}(t)$ in Eq.~(\ref{eq:JnT}),  in recoil units ($k_{\rm recoil}=2\pi/L$) for an initial zero momentum state.  Thin solid line is the current average per cycle $\mathcal{I}_m$ in Eq. (\ref{eq:Im}) obtained with a numerical integration of the 
Schr\"odinger equation for $M=5$, $L=41$, $J/\omega=1.0398$, $V/\omega=0.1$, $\phi/\pi=0.2579$ and $\alpha=1.2$. The driving frequency $\omega$ is tuned to the resonance condition in Eq.(\ref{eq:res}). 
Dashed line corresponds to the current of the effective 3-level system in Eq.~(\ref{eq:T}) and Eq.~(\ref{eq:Im3}). The average current attains a maximum $M$. b) Numerical calculation of $\mathcal{I}(t)$ for same parameters except for solid line  $V/\omega=0.1$, $\phi/\pi=0.2579$; dashed line $V/\omega=10^{-3}\sqrt{10}$,  $\phi/\pi=-0.2579$; dot-dashed line  $V/\omega=10^{-4}$, $\phi/\pi=0.2579$. Note that the time of peak average current scales with  $V^{-2} \omega$. }\label{fig:res1}
\end{figure}
%%%%%%%%5
We consider a driven system of non-interacting bosons with $H(t)=H_0+V(t)$ in a lattice of $L$ sites with periodic boundary conditions \cite{lattice} and
\begin{eqnarray}
H_0 &=&-J\sum_{l=1}^{L} \left|l \right> \left<l+1 \right| + \ket{l+1}\bra{l},
\label{Eq:Ham} \\
V(l,t)&=& V \sin(\omega t) \left[ \sin (\frac{M 2\pi l}{L})+\alpha \sin(\frac{M 4 \pi l}{L}+\phi) \right], \label{eq:v}
\end{eqnarray}
with $M$ integer, where $J$ is the tunneling probability and $\ket{l}$ represents the state of a boson located on site $l$. 
The eigenenergies of $H_0$ are $E_k=-2 J \cos(2 \pi k/L)$ with integer $k=\left[-k_{\rm max},k_{\rm max}\right]$ where $k_{\rm max}=(L-1)/2$ for $L$ odd and 
the corresponding momentum eigenvectors $\langle l \ket{k}=e^{i\frac{ k 2 \pi l}{L}}/\sqrt{L}$ are degenerate for $\pm k$. For convenience one can 
introduce the basis $\langle l \ket{s_{k}}=\sqrt{2/L} \cos(k 2 \pi l /L)$ and $\langle l \ket{a_{k}}=\sqrt{2/L} \sin(k  2 \pi l/L)$ which 
are symmetric or antisymmetric under the inversion of $k$. %Note that this symmetry operation conmutes with $H_0$ but not with $H(t)$ and it is only used to label the basis. 
We add a time and spatially modulated periodic function $V(l,t)$ with frequency $\omega=2\pi/T$ tuned to the $M$-dependent resonant condition  \cite{Sols, creff_sols2}
\begin{equation}
2\omega =E_{2M}-E_0= 2J (1-\cos(4 \pi M/L))  \label{eq:res}
\end{equation} and consider the zero momentum 
$\ket{0}$ as initial state. Note that in contrast to \cite{Sols, creff_sols2} we add the parameters $M$ and $\phi$ to the driving potential in Eq.(\ref{eq:v}).
Parameter $M \leq k_{\rm max}/2$  allows for the coupling of the initial state to very high momentum states $\ket{M}$ and $\ket{2M}$, 
and the parameter $\phi$, key to our model, allows for the coupling between the $\ket{s_{k}}$ and $\ket{a_{k}}$ basis states.

Our choice of a two-harmonic spatial potential times a monochromatic time-dependent potential implies that symmetry $S_t$ (labelled $S_2$ in \cite{Denisov2007}) 
is not broken for $t_s=\pi/(2\omega)$. Therefore $\av{p}_{jj}=0$ and no asymptotic current is possible for our system.  
The average current generated after $n$ cycles arises only from crossed terms between the cyclic states. Our aim is to maximise the average current during any experimental 
time $t_e$.
The key ingredients are to keep few terms in the sum in Eq.(\ref{eq:JnT}), with small exponents and relevant prefactors. The two first are achieved by tuning the resonance in Eq. (\ref{eq:res}) with weak driving $V/J<1$. 
We show that the prefactors can be successfully optimized if the 
quasiresonant cyclic states that have non-zero projection into the initial zero momentum state mix the symmetric and antisymmetric momentum states, which is obtained for $\phi \neq l \frac{\pi}{2}$ for $l$ integer. 
For $\phi=l\pi$, accidental degeneracies could in principle allow to obtain a small non asymptotic current for some specific parameters and a particular value of the coupling \cite{reply}. 
In contrast, we show that for $\phi \neq l \pi/2$ one can tailor an interference between two paths of the same perturbative order and find optimal parameters $M$, $\phi$ and $\alpha$ for any $L$ and $J$ set by $H_0$, to obtain an average current that can be tuned up to near optimal value $\mathcal{I}(t_e) \simeq M \simeq k_{\rm max}/4$, for a time interval $[0,t_e]$ where $t_e$ can be independently tuned by adjusting the driving strength $V/J$.

We show in fig \ref{fig:res1}a) the average current  $\mathcal{I}$
and the oscillating average current per cycle $\mathcal{I}_{m}$ as a function of the number of cycles.
We note that the average current achieves a maximum $M$ in recoil units and, as expected, vanishes for long times. 
We observe in fig \ref{fig:res1}b) that the current changes direction with a sign change in $\phi$ and $t_e$ scales with $V^{-2}\omega$. For the weak driving strength used here the current is nearly zero ($\mathcal{I} \leq 10^{-3}$) for $\phi=0$.

\begin{figure}
\centering
 \epsfig{file=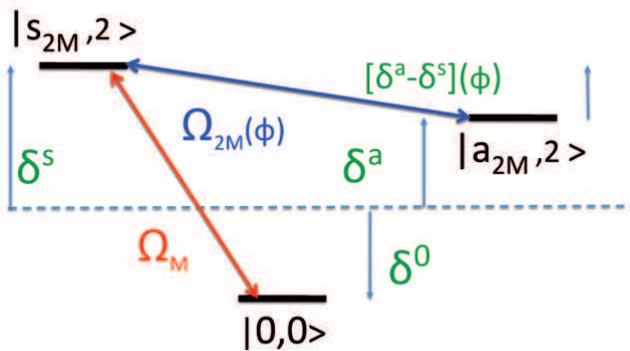, width=\linewidth}
\caption{(color online) Sketch of the second order process in Eq.~(\ref{eq:T}) that takes place for weak coupling $V/J<1$ between the three resonant Floquet states. 
Dashed horizontal line shows the resonance quasienergy for zero driving strength and $\delta$ indicates the quasienergy shifts induced by the potential in Eq. (\ref{eq:v}). Arrows represent the couplings $\Omega$ between the Floquet basis states induced by the driving.
The optimal current shown in Fig. \ref{fig:res1} is 
obtained when the mixing between the symmetric and antisymmetric states is maximized and both couplings in this diagram have the same weight.   
}\label{fig:sketch}
\end{figure}

The previous results are obtained from a full numerical calculation. In order to further understand the effect we follow the methodology of \cite{creff_sols2} and develop an approximate perturbative model. We find that for our potential this simplified model also explains the main features and gives very accurate results for low driving. Close to the resonance Eq. (\ref{eq:res}) and for weak driving  $V/J < 1$  the dynamics of the system involve
only three Floquet basis states $\{\ket{s_{2M}, 2} , \ket{0, 0}, \ket{a_{2M}, 2} \}$,
where $\ket{j}=\ket{k,n}$ with $\langle t \ket{n}=e^{-i n\omega t}$. We apply time-independent perturbation theory in Floquet space, using the $\mathcal{T}$-matrix approach $\mathcal{T}(\epsilon)=V+VG_0(\epsilon)\mathcal{T}$ \cite{tmatrix},
where $G_0(\epsilon)=\sum_j\frac{\ket{j}\bra{j}}{\epsilon-\varepsilon^0_j}$  and $\varepsilon^0_j \equiv E_k-n \omega$. 
Around the ground state quasienergy $\epsilon=\varepsilon^0_0$, the first non-zero term connecting the three states is given by the second order in the expansion $\mathcal{T}(\varepsilon^0_0)\simeq V G_0 (\varepsilon^0_0)V$ which 
 reduces to
\begin{eqnarray}
\small
\mathcal{T} \simeq \frac{V^2}{4} \left( \begin{array}{ccc} \delta^s (\alpha,\phi) & \Omega_{M} &  \Omega_{2M} ( \alpha, \phi)  \\
 \Omega_{M} &\delta^0 (\alpha) & 0  \\   \Omega^\ast_{2M} ( \alpha, \phi)   & 0 & \delta^a (\alpha, \phi) \end{array}\right).
\label{eq:T}
\end{eqnarray}
A sketch of the relevant processes is depicted in Fig.\ref{fig:sketch} and the exact values of the matrix elements, inverse of quasienergy differences which depend on $M/L$, can be found in the complementary material. 
The energy shifts are $\delta$ and the couplings $\Omega$ that correspond to each part of the potential in Eq.~(\ref{eq:v}) are indicated by the subindex $M$ or $2M$. 
Remarkably, $\Omega_{2M} \propto \alpha^2 \sin(2 \phi)$ and thus the coupling between the symmetric and antisymmetric basis states requires $\phi\neq0$. 
The other effect of $\phi$ is to bring those states closer in energy $\delta^s-\delta^a \propto \alpha^2\cos(2\phi)$. 
The optimal current is obtained when the cyclic states (related to eigenvectors of the above matrix) mix the three basis states on an equal foot, corresponding to the 
two second order processes sketched in Fig. \ref{fig:sketch} being of the same order. Due to the structure of the spectrum, 
this optimal mixing can be reached for $M\leq k_{\rm max}/4$. Then all the matrix elements in Eq. (\ref{eq:T}) are of order $\mathcal{O}\simeq 1/\omega$ and the quasienergies $\varepsilon_j \simeq V^2/\omega$, leading to the time scale of the dynamics shown in Fig. \ref{fig:res1}b).

For an initial state $\ket{\psi(0)}=\ket{0}$, the average current at cycle $m$ after evolution with the effective Hamiltonian Eq.~(\ref{eq:T}) reduces to a sum of $3$ oscillatory terms with different frequencies and the same weight 
\begin{eqnarray}\small
 \mathcal{I}_{m}&=& \mathcal{C} (\alpha,\phi)  \sum_{j<j'}\sin(mT \Delta \varepsilon_{jj'})  , \label{eq:Im3} 
\end{eqnarray}
where $\mathcal{C} (\alpha,\phi) = 4 i c_1^* c_2 \av{p}_{12} $ with $p=2 (\ket{s_{2M}}\bra{a_{2M}}+ \ket{a_{2M}}\bra{s_{2M}})$ 
in the reduced basis and $\Delta \varepsilon_{jj'}:= (-1)^{j+j'}(\varepsilon_j-\varepsilon_{j'})$. 
Once the prefactors $c_1^* c_2$ are optimized for any given $M \leq k_{\rm max}/4$, the current is linear with $M$. 
Thus, we can set $M_{\rm opt}$ as the closest integer to  $k_{\rm max}/4$ and obtain a robust near optimum value for the integrated 
current of $\mathcal{I} \simeq M_{opt}$. As shown in figure \ref{fig:res1} (dashed line), the three-mode approximation Eq.~(\ref{eq:Im3}) fits perfectly the exact numerical results.

\begin{figure}
\centering
%\begin{tabular}{cc}
\epsfig{file= 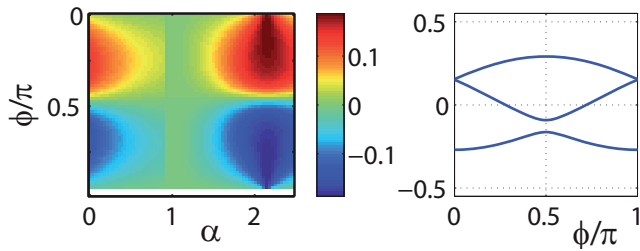,width=\linewidth,clip=} 
%\end{tabular}
\caption{(color online) Left panel: Amplitude $\mathcal{C}/4M$ of the current in Eq.(\ref{eq:Im3}) 
for different values of the potential parameters $\phi$ and $\alpha$ for an initial zero momentum state and $L=41$, $M=5$, $J/\omega=1.0398$. 
Right panel: Eigenenergies of the effective $\mathcal{T}$-matrix in Eq.(\ref{eq:T}) in units of $V^2/(2\omega)$ as a function of $\phi$ for $\alpha=1.2$. For fixed values $\phi$ and $\alpha$ the average current per cycle in Eq. (\ref{eq:Im3}) 
is a sum of sinusoidals with an amplitude $\mathcal{C}$ shown in the left panel and frequencies given by the quasienergy differences.}

\label{fig:parC}
\end{figure}

We show the current amplitude $\mathcal{C}$ in the left panel of figure \ref{fig:parC} for different parameters $\alpha$ and $\phi$ . As explained above, it attains its maximum at  $\alpha_{\rm opt} \simeq 1.2$ when both terms of the driving potential have the same weight, it is periodic in $\phi$ with $\pi$ periodicity and has vanishing values for $\phi= l \pi/2$ with $l$ integer. For $\phi_{\rm opt}\simeq \pm \pi/4$ one can see in the right panel of  fig. \ref{fig:parC} that the quasienergies become equidistant and thus there is only one relevant energy scale $\Delta \varepsilon_{f}=0.205 V^2/\omega$ in the sum in Eq. (\ref{eq:Im3}) while the other sinusoidals oscillate with half this frequency. One can then average $\mathcal{I}_m$ over different periods to obtain $\mathcal{I}$ which achieves its maximum $\mathcal{I}(t_{\rm e}=n_{\rm e} T)\simeq M$ after $n_{\rm e}=1.81(V/\omega)^{-2}$ cycles as shown in fig. \ref{fig:res1} b). 

\begin{figure}
%\begin{tabular}{c}
%\epsfig{file= fig5.eps,width=0.9\linewidth,clip=} \\
\epsfig{file= 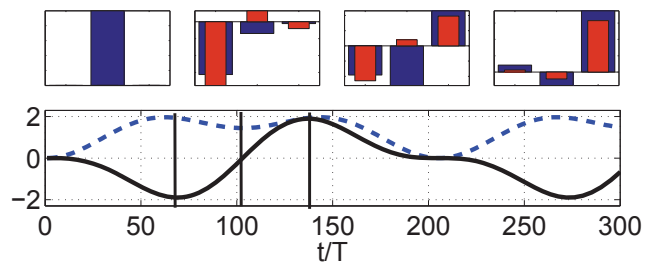,width=\linewidth,clip=}
%\end{tabular}
\caption{Upper panel: Real (thick bar) and imaginary part (thin bar) of the $\{\bra{-2M}, \bra{0},\bra{2M}\} \ket{\psi (t)}$ at $t=0$ and at the times showed by vertical lines in the lower panel. 
Lower panel: Average current per cycle $\mathcal{I}_m/M$ in recoil units and average tunneling energy per cycle (dashed line) in units of $1/\omega$ as a function of time.
Same parameters as in fig. \ref{fig:res1} a).}\label{fig:resK}
\end{figure}

In the context of cold atoms it may be of interest not only the generation of a current from an initial zero momentum state, 
but also the control of the quantum state of the system.  We plot in fig. \ref{fig:resK} the particle state in the momentum basis and the average current per cycle and the average kinetic energy $\int_0^T dt \langle H_0\rangle/T$. We observe that the zero momentum state can be indeed converted into an almost pure momentum state $\ket{\pm 2M}$. 
One could then switch off the driving, thus breaking the time reversal symmetry $S_t$, and use this scheme to generate an asymptotic current. This is an example of the high controllability of our system. 

\begin{figure}
\epsfig{file=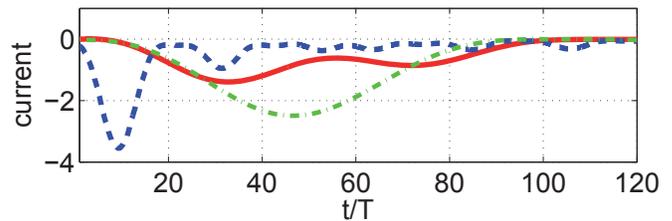,width=\linewidth,clip=}
\caption{ Average current in recoil units in Eq.~(\ref{eq:JnT}) as a function of time for an initial zero momentum state. 
Parameters $M=5$ , $L=41$, $J=1.0398$, $\phi/\pi=0.2579$ and $\alpha=1.2$. Dashed line corresponds to $V=0.2$ and $\omega=1.01$, thick line $V=0.5$ and $\omega=1$, dash-dot line: $V=0.2$ and $\omega=1$ with an initial state $ \langle k | \psi (0) \rangle\sim\exp(-k^2/1.5^2)$ }\label{fig:res2}
\end{figure}

Finally, let us analyze the feasibility of the model. We can summarise our findings in a simple recipe. $H_0$ sets  the energy scale $J$ and the length $L$ of the system. 
We can obtain a final average current $\mathcal{I}$ of nearly $(L-1)/8$ in recoil units for a given time interval $[0,t_{\rm e}]$
if one tunes a driving potential in Eq.~(\ref{eq:v}) with parameters $\omega$ from Eq.~(\ref{eq:res}), $M_{\rm opt}$, $\alpha_{\rm opt}$, $\phi_{\rm opt}$  
defined above and $V=\sqrt{11.37 J (1-\cos(4 \pi M_{\rm opt}/L))/t_{\rm e}}$ with the constraint that $V/J\leq 1$. 
We show in fig \ref{fig:parC} that small changes in $\alpha$ and $\phi$ around optimal values only slightly affect the current. Smaller $M$ would reduce the maximum average current attained and require adjustment of the resonance condition in Eq.(\ref{eq:res}). Thus the only actual requirements 
of our model are that the system is tuned to resonance and that the driving field is weakly coupled. To show the robustness of the method we plot in figure \ref{fig:res2} the average  current when the system is not perfectly tuned to resonance and when the driving field is stronger. We find that couplings up to $V/\omega=0.5$ and errors of $1\%$ in the resonant frequency still give rise to high particle currents. Note that due to the resonance condition at low driving the effect is highly selective. If the initial state is a narrow wavepacket centered at $k=0$ only this component is mixed to $k=2M$, whereas other $k \leq M$ components remain uncoupled. The reduced average current will be just proportional to the weight of $k=0$ state, see fig. \ref{fig:res2}.

%**********************  fin edicion julio 1 marzo 

We have presented a model system where one can obtain currents with amplitudes orders of magnitude larger than those 
observed in recent experiments with coherent ratchet currents \cite{Salger_2009}. 
The oscillation period of the current can be controlled by the amplitude of the driving potential and in particular, by decreasing the driving strength 
one obtains currents which do not decay during the lifetime of the experiment. 
This effect is obtained with a potential which does not break the time-reversal symmetry and is due to 
crossed terms between the cyclic states of the system. 
The proposed scheme requires that the system is tuned to resonance and that the driving potential is weakly coupled such that only a
few cyclic states are involved in the dynamics. We have checked the robustness of the method by tuning the system out of resonance and 
increasing the coupling strength. Furthermore, we have shown that it is possible to control the quantum state and the amount of kinetic energy in the system,
 using the proposed scheme to convert a zero momentum state into a state with high finite momentum, and viceversa.

We acknowledge support from the Spanish MICINN through projects FIS2007-61686, FIS2009-07277, FIS2010-18799 and the Ram\'on y Cajal programme.

%%%%%%%%%%%%%%%%%%%555

\end{document}